 \documentclass[final,5p,times,twocolumn]{elsarticle}

\usepackage{amssymb}
\usepackage{amsmath}
\usepackage{float}
\usepackage[none]{hyphenat}
\journal{}

\begin{document}

\begin{frontmatter}



\title{Generation of Granular Deposition Interfaces using conditional Generative Adversarial Network (cGAN)  }


\author[label1]{Seyed Feyzelloh Ghavami Mirmahalle} 

\affiliation[label1]{organization={Department of Physics, Institute for Advanced Studies in Basic Sciences (IASBS)},
            city={Zanjan},
            postcode={45137-66731}, 
            country={Iran}}
\author[label1]{Seyed Ehsan Nedaaee Oskoee} 
\author[label2,label1]{Maniya Maleki} 
\affiliation[label2]{organization={Department of physics, University of Northern Iowa},
            city={Cedar Falls},
            postcode={IA 50614},
            country={USA}}

\begin{abstract}
This work aims at generating 1D interface profiles of granular deposition by a conditional generative adversarial network (cGAN). Our cGAN model employs a U-Net generator and a ResNet discriminator that, in competition with each other, produce granular interfaces. The network is trained on dynamic simulation data from the LAMMPS granular package. Different fluids (water, acetone, and hexane) were used for the medium of the deposition cell to check the model performance in different growing conditions. The same model with the same hyperparameters was trained on data from different media separately. The ML-generated interfaces are compared with those of dynamic simulations, and a large number of interfaces are then produced to obtain more stable statistical properties of granular deposition. This way, the computationally extensive molecular dynamics simulation is substituted by the AI model. The statistical trend of interface growth is diagrammed, and the generated interfaces are also analyzed in terms of statistical features.

\end{abstract}




\end{frontmatter}



\section{Introduction}
\label{Introduction}

Surface growth is a fascinating and widely studied phenomenon, appearing in diverse natural and artificial processes such as fire-front propagation, bacterial colony expansion, crystal formation, vapor deposition, and material erosion \cite{Barabasi, Wang}. Granular deposition—such as the settling of seeds, grains, or sand—provides a prototypical example of dynamic surface growth \cite{Wang, KURNAZ93, Sardari}.

One of the simplest theoretical descriptions of surface growth is the \emph{random deposition} model, in which particles fall vertically and attach to the highest particle at the chosen site, without lateral movement. A refinement of this model is \emph{random deposition with surface relaxation}, where particles fall on randomly selected sites and then relax to the lowest neighboring height. A more sophisticated approach is the \emph{Kardar–Parisi–Zhang} (KPZ) model, which incorporates lateral growth and overhang formation \cite{Barabasi, kardar.Parisi.Zhang}.

To study the surface growths, both lattice-based and off-lattice simulations are widely used. Lattice-based models are computationally efficient but less realistic for granular systems, whereas off-lattice simulations, which consider inter-particle forces and boundary interactions, provide greater realism at the cost of higher computational cost. To study such systems, scaling concepts and critical exponents are standard analytical tools. These methods have been successfully applied to a variety of systems, including thin-film growth \cite{MENDOZA.RINCON, NehaSharma}, rock surface morphology \cite{KOEHN.Köehler}, colloidal systems \cite{Oliveira.Reis}, and granular deposition \cite{Ghavami2025}.

In recent years, machine learning has emerged as a powerful tool for modeling complex physical systems. For granular materials, applications include predicting force chains in solids \cite{RituparnoMandal}, estimating force networks using graph neural networks (GNNs) \cite{Cheng}, modeling granular dynamics with recurrent neural networks (RNNs) \cite{QU2021103046}, and characterizing granular behavior using deep neural networks (DNNs) \cite{YE2022109437}. 

T. Song et al \cite{Tianshu} used a CNN model to produce interfaces that originate from the KPZ equation. In their work, the noise term in the KPZ equation is first removed, and the CNN model is trained on the interfaces made by the linear and nonlinear terms of the KPZ equation. And when the model is trained, it is used to generate interfaces, and then different types of noises are added to the generated interfaces. 

Generative Adversarial Networks (GANs) \cite{Goodfellow} provide another machine learning framework well-suited to modeling complex spatial structures. A GAN consists of two competing models: a \emph{generator} $G$, which learns to capture the data distribution, and a \emph{discriminator} $D$, which distinguishes between real and generated samples. Since their introduction, many GAN variants have been developed, offering improved performance or domain-specific capabilities. In \emph{conditional} GANs (cGANs) \cite{Mirza}, $G$ and/or $D$ are conditioned on additional information. One notable application is \emph{image-to-image translation}, where the generator is conditioned on an input image. Using a U-Net-based generator, Isola \emph{et~al.} \cite{Isola} demonstrated high-quality translations between domains such as day-to-night, grayscale-to-color, and aerial-to-map imagery.

In this study, we use a conditional generative adversarial network (cGAN) to learn the mapping from lattice-based random deposition to the interface profile of the off-lattice dynamic simulation. The model architecture consists of a U-Net-based generator and a ResNet-based discriminator. This approach aims not only to substitute computationally extensive dynamic simulation of interface generation by an AI model using the random deposition data but also to investigate the power of generative machine learning models in uncovering complex relationships in physical systems to make new interfaces that include the statistical physics of the interface growth.

\section{Methodology}

\subsection{Machine Learning Model}
\label{Machine Learning Model}

In this work, we employ a conditional generative adversarial network (cGAN) in which the generator is based on the 1D-input U-Net architecture and the discriminator is implemented as a 1D-input ResNet.

\medskip
\noindent{Generator architecture:}
The generator is a U-Net encoder–decoder with skip connections linking corresponding layers in the encoder and decoder paths. The encoder consists of successive one-dimensional convolutional and downsampling layers that extract multi-scale feature representations from the input sequence. The decoder symmetrically reconstructs the output through one-dimensional upsampling and convolutional layers, while incorporating encoder features via skip connections. The final output layer applies a one-dimensional convolution with a sigmoid activation function to produce a 932-element output vector.

\medskip
\noindent{Discriminator architecture:} 
The discriminator is a 1D ResNet model consisting of an initial convolutional layer, followed by two residual blocks, three fully connected layers, and a final output neuron for binary classification (real/fake). We employ a single-output discriminator to focus on preserving more global than local features. A conditioned input representing the simulation time is concatenated at the output of the residual blocks, before the fully connected layers. This way, the discriminator understands the time of deposition, while the generator can understand it by the oscillations in height values in the input data.

\medskip
\noindent{Training setup:} 
The model is implemented in a TensorFlow environment with Adamax optimizer for both the generator and discriminator. The batch sizes were 240, the same as the deposition time of each simulation, and 40 epochs of training were used for training.

\begin{figure*}[h!]
\centerline{\includegraphics[width=2.2\columnwidth, keepaspectratio]{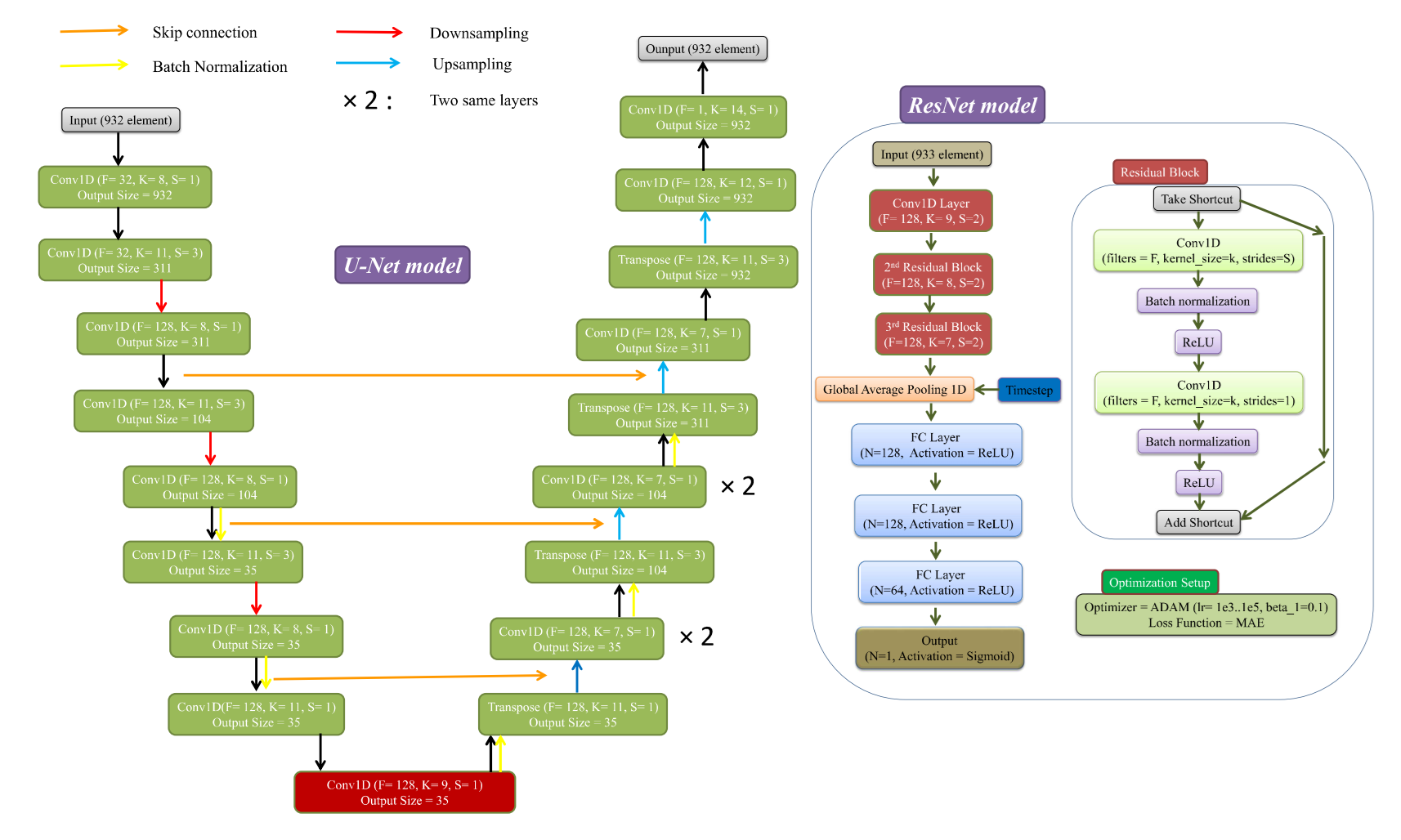}}
\caption{Architecture of the proposed cGAN. Left: U-Net-based generator. Right: ResNet-based discriminator.}
\label{Model}
\end{figure*}

The generator is trained using the mean absolute error (MAE) loss, while the discriminator uses binary cross-entropy (BCE) loss. The model structure is shown in Figure~\ref {Model}.

\subsection{Data Preparation}
\label{Data Preparation}

The data were obtained from the simulations using the granular package in LAMMPS, which is specialized for discrete element method (DEM) simulations of granular materials. Grain–grain and grain–wall interactions were modeled using the Hertzian contact model.

\medskip
The simulations were conducted in two dimensions within a rectangular box of $10\ \mathrm{cm} \times 12\ \mathrm{cm}$. Grains with parameters of silicon spheres were initially placed at random positions between heights of $10$ and $11\ \mathrm{cm}$. 60 grains were created in each simulation loop, where an effective gravitational acceleration considering $g = 9.8\ \mathrm{m/s^2}$ and the density of grain and liquid, acted on all of them. The viscous drag force was modeled using the physical properties of medium liquids and grains. The simulations were carried out for three different fluids in the deposition cell: water, acetone, and hexane.
The simulations were performed in 104-second time steps in LAMMPS. The images for analysis were taken after each set of grains had well settled, which took different times for different media (a few seconds for hexane, up to 30 seconds for water). These images give the timeframes of our data.

For random deposition interfaces, we considered groups of grains falling from the top of the cell, and their $x$-coordinates (in terms of pixels) were extracted from images generated by LAMMPS. These coordinate values for each point of the horizontal axis were then added over time by each new group according to their $x$-position. It provided the random deposition profile at each deposition time. It included information on the number of grains that fell at each point on the horizontal axis over time.

\begin{figure}[h!]
\centerline{\includegraphics[width=1.1\columnwidth, keepaspectratio]{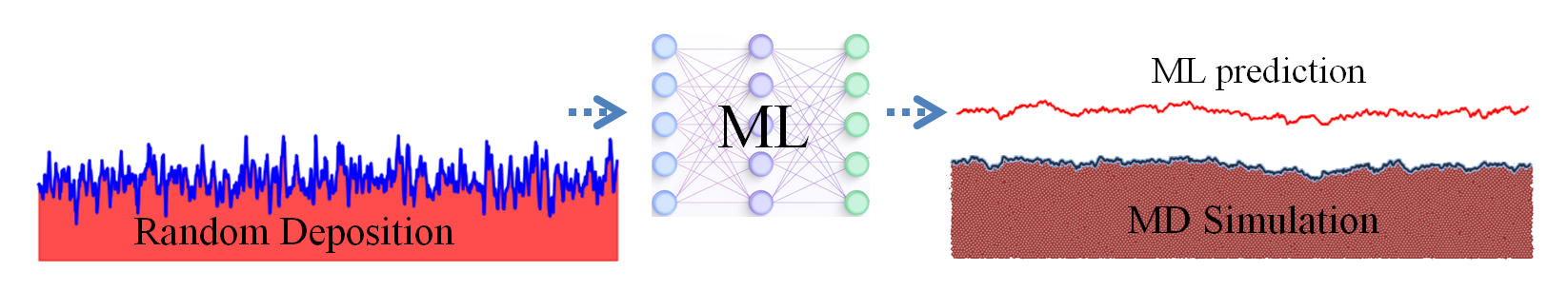}}
\caption{One simulation snapshot of deposition in the water medium. The noisy curve in the left image represents the random deposition interface at one time frame. At right, the interface generated by dynamic simulation for the same time frame is shown, and the corresponding model prediction of the interface is shown above it in red color.}

\label{setup}
\end{figure}

In total, 246 groups were deposited, resulting in 246 time frames. However, the first 6 groups are excluded from the dataset, as their behavior occurs on a small timescale and is nearly identical across all simulation runs. Therefore, 240 timesteps of both random deposition and dynamic deposition profiles, each repeated 50 times, are used as model inputs and outputs for each of the three different liquids. 

Three different media were selected for training in the model, as the medium would affect the global behavior of interface growth. For each medium, we had 50 dynamic simulations and 50 corresponding random depositions. 450 random depositions, each with 240 time frames, were also performed without dynamic simulations. By doing so, we could predict corresponding dynamic interfaces in an unsupervised manner. 

The growing interfaces obtained from random deposition simulations were used as inputs to the model. For each case, a corresponding dynamic deposition simulation provided the target output interface. In total, 50 random deposition simulations were performed, each paired with one dynamic deposition simulation.

42 simulations of each medium were used for training, 4 for validation, and 4 for testing. Each simulation contained 240 surface profiles, resulting in $42 \times 240$ training samples. During training, the samples were not shuffled, and the model was taking each simulation run as a batch of 240 arrays. The validation and test samples were also maintained in their original order, allowing for comparison of the model predictions with the MD profiles as a function of time.

The generator used the height profile of the random deposition interface as input and produced a predicted interface that resembled the one obtained from dynamic simulations in LAMMPS. The discriminator was trained to distinguish between real interface profiles (generated through dynamic simulation) and those generated by the U-Net. Each interface in the input and output data was subtracted by its mean before being injected into the model to avoid training by the average height of the interface. 

Thus, the model was trained on input–output pairs from 50 simulations over the time frames, where the input was the interface from random deposition and the output was the corresponding interface from dynamic deposition. After training, the model was applied to predict interfaces from 4 random depositions of test data and 500 random depositions.

{Data augmentation:} 
Considering the deposition process in simulation, horizontally mirroring the interface does not change the statistical properties of the interface. This symmetry was exploited for data augmentation, where each array of random deposition interfaces and its corresponding dynamic deposition interface were mirrored about the vertical axis. The mirrored pairs were then added to the training set as new samples.

\subsection{Scaling Concepts}
\label{Scaling Concepts}

Scaling theory provides a robust framework for characterizing interface growth. It encompasses a broad range of concepts and equations applicable to this field. Here, we focus on the main equations and the primary concepts that can statistically evaluate the outputs of our cGAN model. 

Two fundamental parameters for describing an interface are: 
(1) the mean interface height, and (2) the interface roughness (also called interface width). 

The mean height $\Bar{H}(t)$ evolves according to
\begin{equation}
\Bar{H}(t) = \frac{1}{N} \sum_{i=1}^{N} H(x_i,t),
\end{equation}
where $H(x_i,t)$ is the height of the growing interface at position $x_i$ and time $t$.

The second parameter, the roughness of the interface, is defined as
\begin{equation}
W(L,t) = \left[ \frac{1}{N} \sum_{i=1}^{N} \left( H(x_i,t) - \Bar{H}(t) \right)^2 \right]^{1/2}.
\end{equation}

The plot of $W$ versus $t$ reveals the key features of roughness evolution in interface growth. From this curve, one can identify the \emph{saturation time}, beyond which the roughness no longer increases, as well as the initial growth rate of the roughness \cite{Barabasi}. In this study, we present these roughness plots but do not go into further detail regarding the underlying physics of the simulations.

\section{Results}

In this study, we utilized the capabilities of machine learning to generate new interfaces for granular deposition. The generative model was trained using the interfaces from molecular dynamics simulations to produce new interfaces. These interfaces were related to different media of deposition, different times of deposition, and different random distributions. We use one model with the same hyperparameters to handle all the complexities. The training process was carried out independently for the data from each medium of deposition, but under an identical configuration of the model. 

After training, the model was first used to predict interface profiles on the test dataset. To show the statistical features, we have calculated the average roughness growth of interfaces over time for the three media. For each medium, we have 50 simulations with different random distributions.
Another 50 random depositions, each in 240 time frames, were selected for each medium, and the interfaces were produced by the generator in an unsupervised way with no corresponding dynamic simulations. We used our model to predict a total of 500 simulations for each medium, as increasing the number of simulations and averaging them reduces noise, resulting in more accurate diagrams of surface deposition.

\subsection{Water Medium}
\label{Water Medium}

\begin{figure}[H]
\centerline{\includegraphics[width=1.0\columnwidth, keepaspectratio]{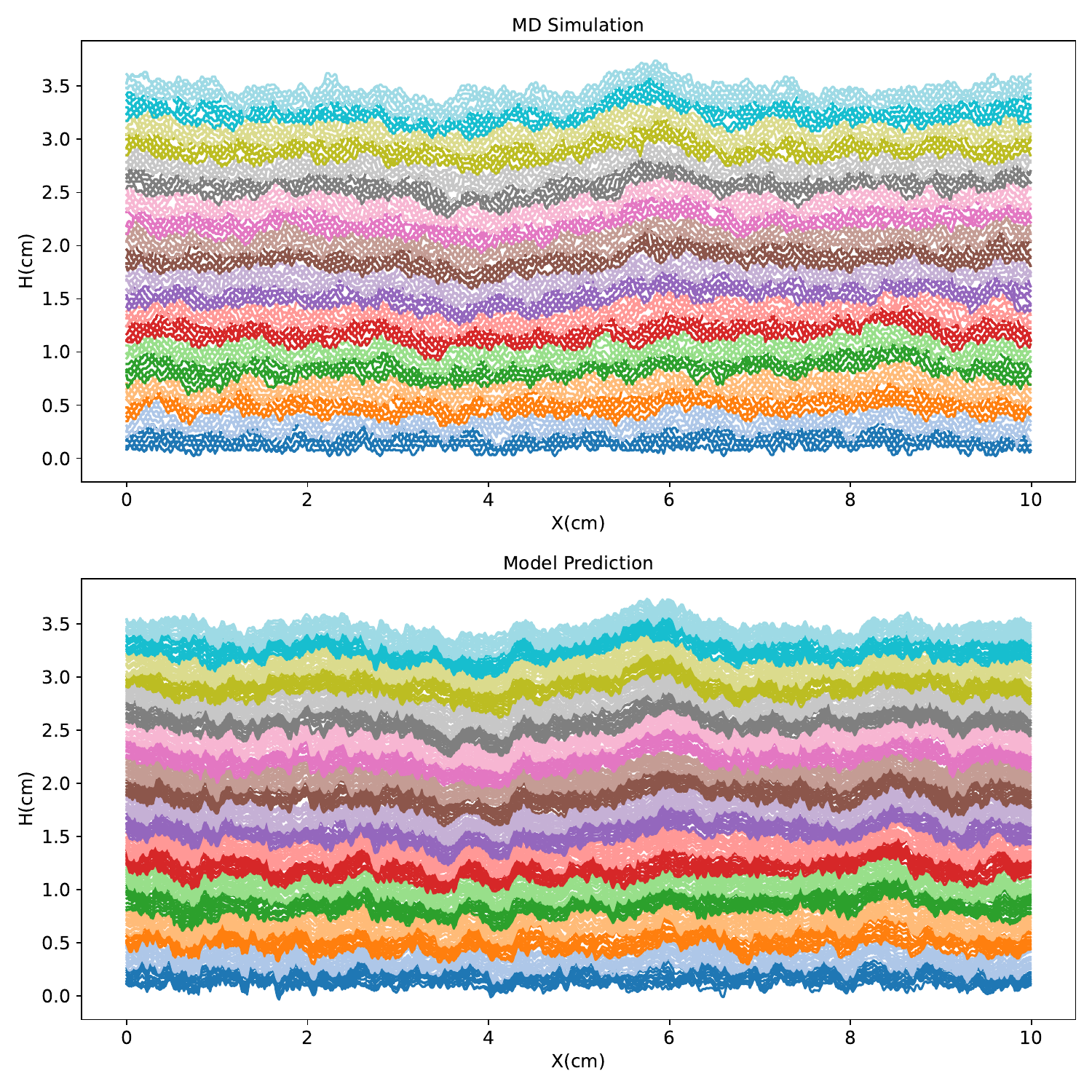}}
\caption{The upper image shows the growth of 240 interfaces during time for one dynamic simulation in water medium, and the lower image shows 240 of the corresponding model predictions in the test data. Different colors only represent different times of deposition.}
\label{all-water}
\end{figure}

To visualize how the generator produces results over time, all interfaces of a molecular dynamics simulation run and the corresponding model predictions at 240 different times for the water in the experimental data are shown in~\ref{all-water}. 

\begin{figure}[H]
\centerline{\includegraphics[width=1.0\columnwidth, keepaspectratio]{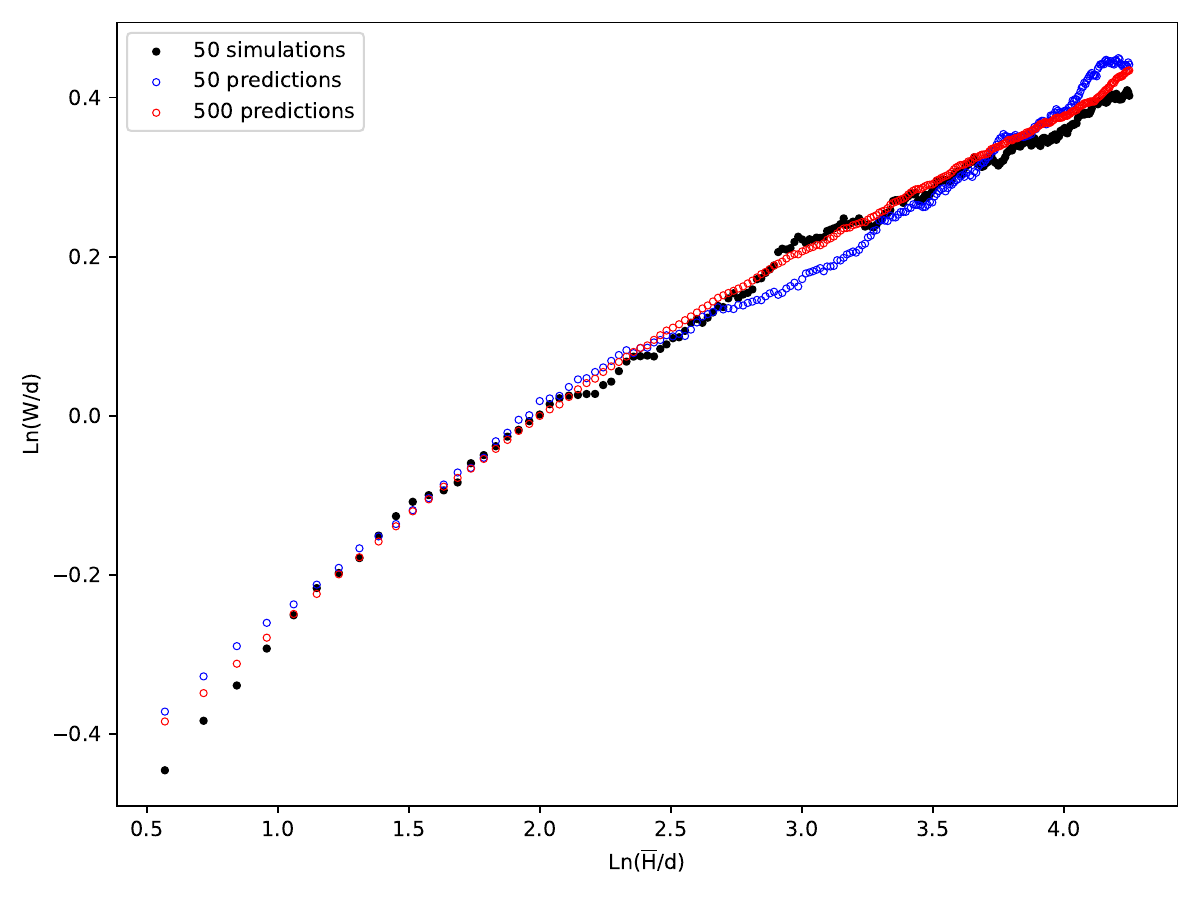}}
\caption{Roughness diagram of water medium for 50 dynamic simulations, 50 predictions of the model, and 500 predictions of the model.}
\label{roughness-water}
\end{figure}

The diagram of roughness in terms of averaged height for water medium deposition is shown in Figure~\ref{roughness-water}. 

\subsection{Acetone Medium}
\label{Acetone Medium}

\begin{figure}[H]
\centerline{\includegraphics[width=1.0\columnwidth, keepaspectratio]{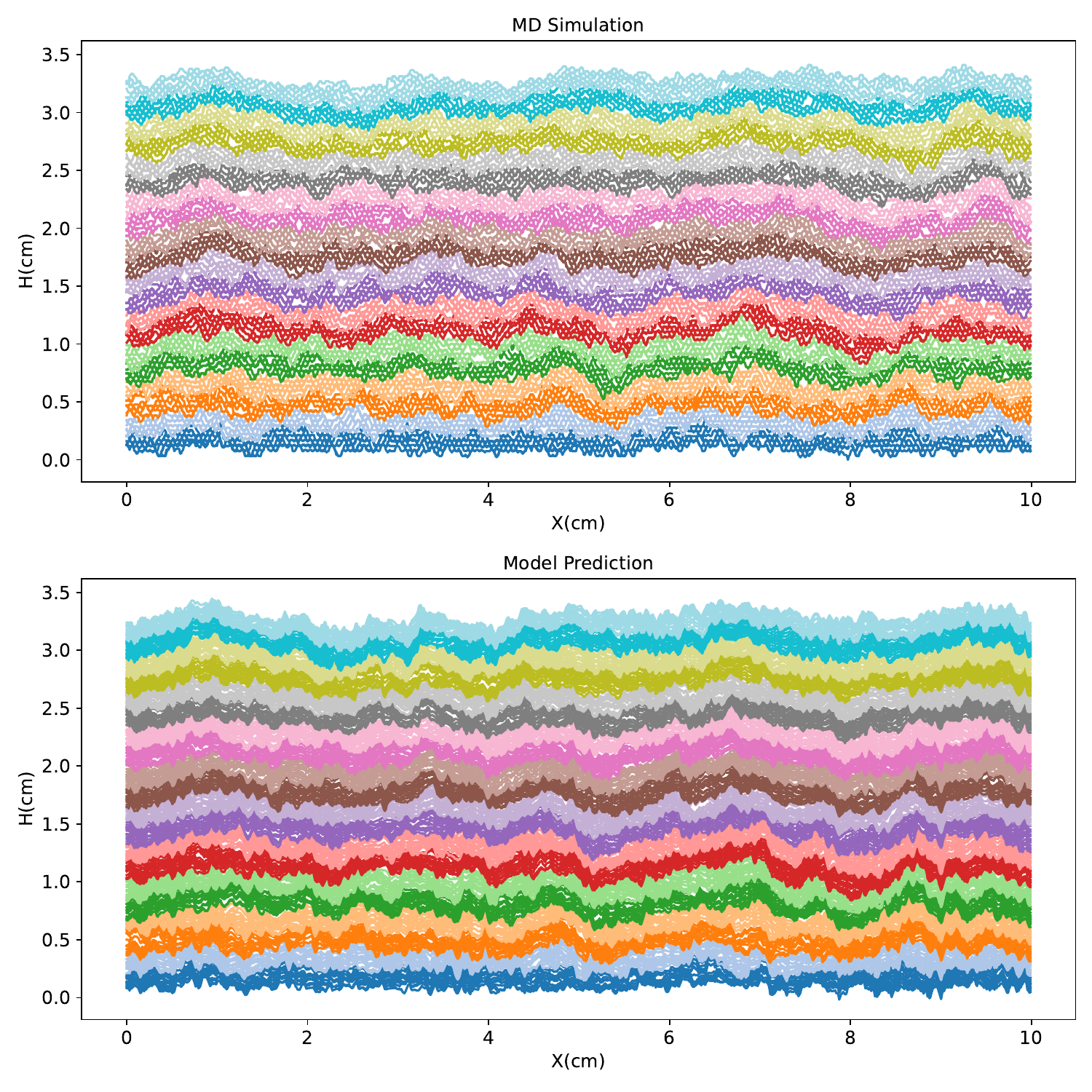}}
\caption{The upper image shows the growth of 240 interfaces during time for one dynamic simulation in acetone medium, and the lower image shows 240 of the corresponding model predictions in the test data. Different colors only represent different times of deposition.}
\label{all-acetone}
\end{figure}

All the interfaces of one running MD simulation and their corresponding model predictions at 240 times for the acetone medium in the test data are shown in Figure~\ref{all-acetone}. 

\begin{figure}[H]
\centerline{\includegraphics[width=1.0\columnwidth, keepaspectratio]{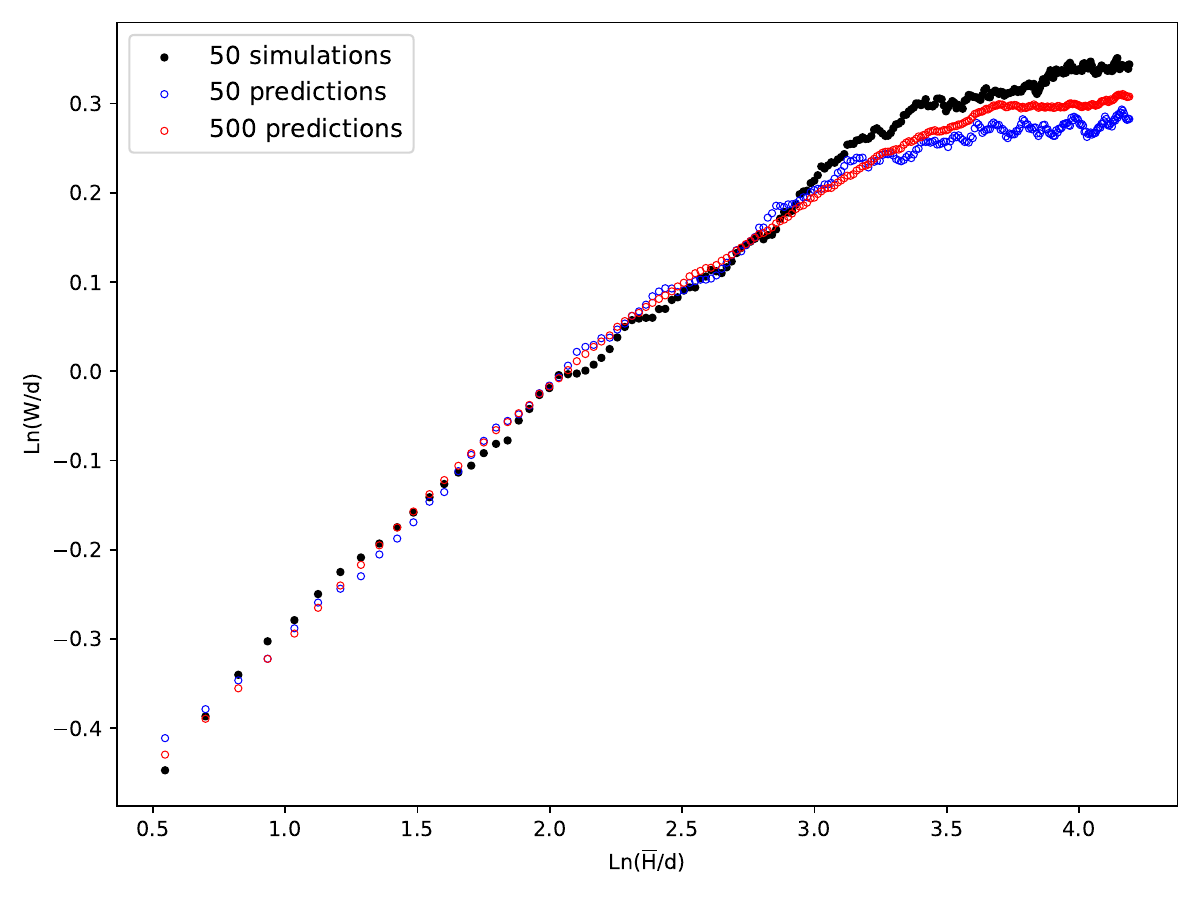}}
\caption{Roughness diagram of acetone medium for 50 dynamic simulations, 50 predictions of the model, and 500 predictions of the model.}
\label{roughness-acetone}
\end{figure}

The diagram of roughness in terms of averaged height for acetone medium deposition is shown in Figure~\ref{roughness-acetone}. 

\subsection{Hexane Medium}
\label{Hexane Medium}

\begin{figure}[H]
\centerline{\includegraphics[width=1.0\columnwidth, keepaspectratio]{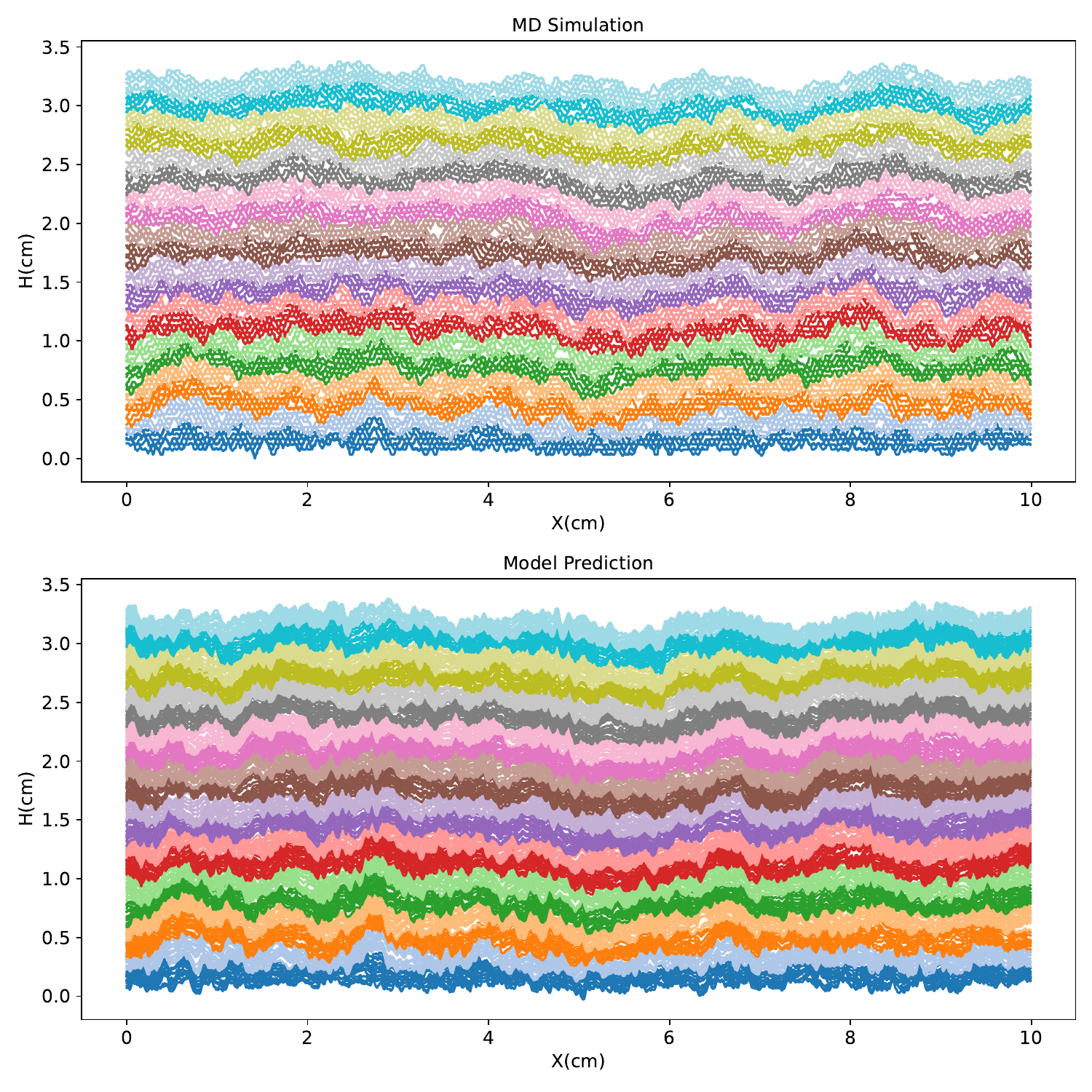}}
\caption{The upper image shows the growth of 240 interfaces during time for one dynamic simulation in hexane medium, and the lower image shows 240 of the corresponding model predictions in the test data. Different colors only represent different times of deposition.}
\label{all-hexane}
\end{figure}

All the interfaces of one MD simulation run and their corresponding model predictions at 240 times for the acetone in the test data are shown in Figure~\ref{all-hexane}.

\begin{figure}[H]
\centerline{\includegraphics[width=1.0\columnwidth, keepaspectratio]{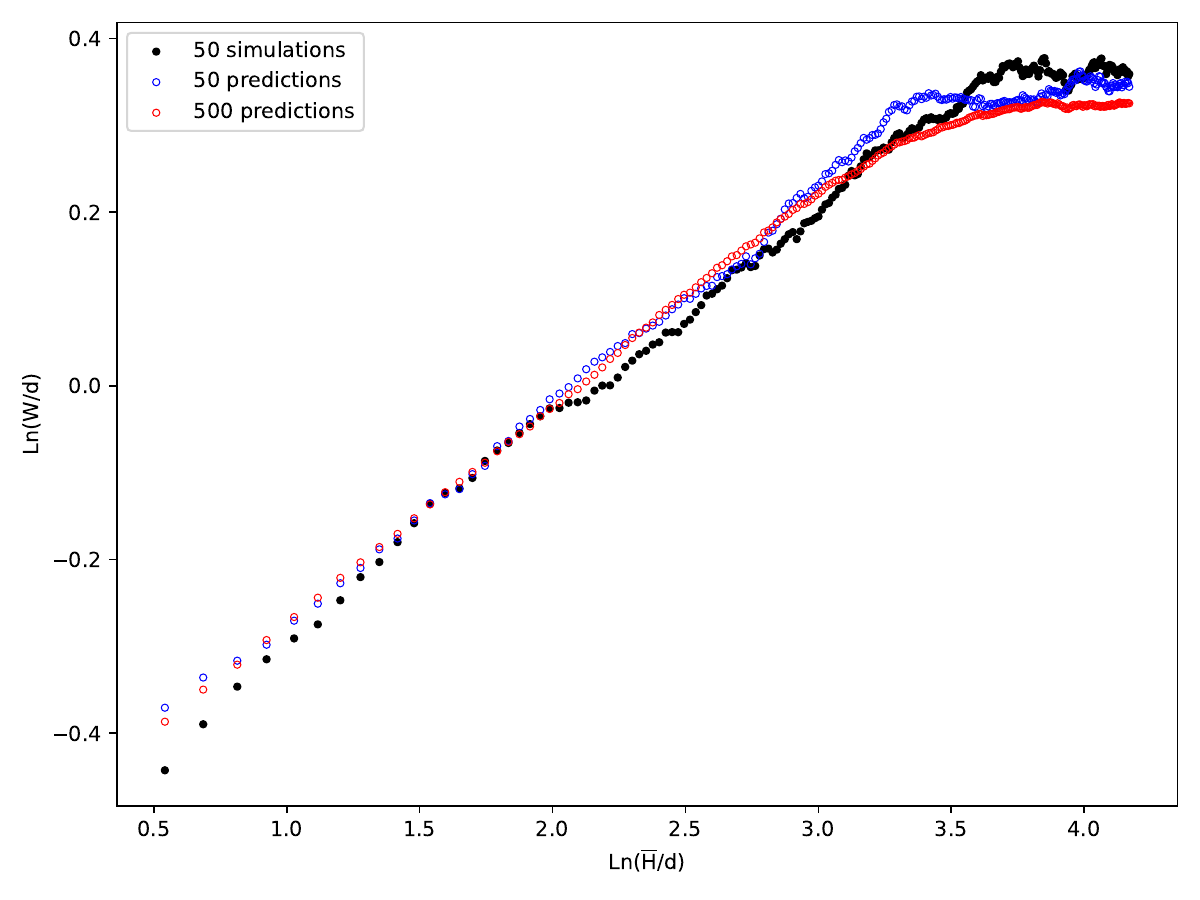}}
\caption{Roughness diagram of hexane medium for 50 dynamic simulations, 50 predictions of the model, and 500 predictions of the model.}
\label{roughness-hexane}
\end{figure}

The diagram of roughness in terms of averaged height for hexane medium deposition is shown in Figure~\ref{roughness-hexane}. 

\section{Discussion}

This study examines the use of a machine learning model to help reproduce interfaces of granular deposition. The dynamic simulations were performed with different fluids as media of deposition. As a result, the simulations were conducted with different gravity forces and kinetic energies of grains, which resulted in different interface growth types. In the water medium, the fluid has a density close to that of the grains and a high viscosity compared to other media; therefore, there is minimal movement of the grains. The medium fluid of hexane had the highest density difference, resulting in the highest effective gravity force on the grains, and the lowest viscosity, which allowed the grains to have higher kinetic energy and greater freedom of movement. While the number of deposited particles is the same for the three fluids, it can be seen that water does not exhibit saturation, which clearly implies a different state of growth among the fluids. The prediction of interfaces was performed by the same hyperparameters and individually for each medium. Roughness diagrams were plotted to verify whether the cGAN model generates interfaces that exhibit the statistical properties of the depositions. 

The results demonstrate that a conditional generative model can successfully learn the complex mapping from a random-deposition interface of falling grains to the dynamically simulated interface produced by off-lattice granular simulations. The U-Net–ResNet cGAN produces height profiles whose large-scale shape and roughness evolution are comparable to those observed in the LAMMPS simulation data, as was also observed in the test data. In all three fluid regimes (water, acetone, and hexane), the predicted interfaces reproduce the qualitative features of the dynamically simulated surfaces, as seen through their roughness diagrams. This indicates that the generator has effectively learned the underlying dynamics and stochastic variability of the deposition process. While the discriminator is given by the timestep of deposition, the results show that the generator has been capable of distinguishing the time of deposition from the changes in random deposition heights. 

To our knowledge, this is the first application of deep generative modeling on surface growth and the first machine learning model on granular deposition. The model of KPZnet, as presented by T. Song et al. \cite{Tianshu}, is limited to the KPZ equation without incorporating noise, where the noise is added to their outputs outside the neural network. While in our work on granular deposition, the cGAN model handles the noise itself.

While using the model in our work does not yield as accurate results as dynamic simulations, it provides significant computational efficiency. In our experiments, generating 500 predicted profiles for each of the 240 time series took less than an hour on a multi-core CPU, whereas it would take months for the full dynamic simulations in the case of the water medium. At the same time, the limitation is that it needs primary simulation to be trained and then reproduce it. 

\section*{Conclusions}

We have developed and evaluated a conditional GAN (cGAN) that uses a U-Net generator and a ResNet discriminator to generate granular interfaces from random deposition interfaces. Trained on LAMMPS simulation data that include Hertzian particle interactions, the model successfully learned the complex surface growth dynamics for different media. The predictions of interface growth for the three media, water, acetone, and hexane, were well performed by the model. The roughness diagrams of the predicted interfaces were similar to those of the dynamically simulated ones when averaged over 50 growths. The diagrams of the averaged 500 growths showed significantly less noise compared to the averaged 50 diagrams, indicating appropriate physical results.


\begin{thebibliography}{99}

\bibitem{Barabasi}
Barabási, A.-L.,
\textit{Fractal concepts in surface growth},
Cambridge University Press, 1996.

\bibitem{Wang}
Wang, C., Xia, H.,
\textit{Scaling properties and height distributions of persisting roughness in the discrete growth models in the presence of the angle of repose},
Chaos, Solitons \& Fractals, 181, 114598, 2024. DOI: 10.1016/j.chaos.2024.114598


\bibitem{KURNAZ93}
Kurnaz, M.L., McCloud, K.V., Maher, J.V.,
\textit{Sedimentation of Glass Beads under the Influence of Gravity},
Fractals, 1(4), 1008–1021, 1993. DOI: 10.1142/S0218348X93001106


\bibitem{Sardari}
Sardari, V., Safari, F., Maleki, M.,
\textit{Dynamics of the surface growth resulted from sedimentation of spheres in a Hele–Shaw cell containing a low-viscosity fluid},
Physics of Fluids, 36(5), 053303, 2024. DOI: 10.1063/5.0200886


\bibitem{kardar.Parisi.Zhang}
Kardar, M., Parisi, G., Zhang, Y.-C.,
\textit{Dynamic scaling of growing interfaces},
Phys. Rev. Lett., 56, 889, 1986.


\bibitem{MENDOZA.RINCON}
Mendoza-Rincón, S., Ospina-Arroyave, M.S., Arias Mateus, D.F., Escobar-Rincón, D., Restrepo-Parra, E.,
\textit{Substrate rotation effect over scaling roughness exponents in Zr thin films grown by GLAD technique},
Applied Surface Science, 559, 149660, 2021. DOI: 10.1016/j.apsusc.2021.149660

\bibitem{NehaSharma}
Neha Sharma and K. Prabakar and S. Dash and A.K. Tyagi,
\textit{Ion beam sputter deposition, Dynamic scaling theory, Growth kinetics, Static and dynamic scaling constant, Scaling phenomenon},
Applied Surface Science, 573, 2014. DOI: 10.1016/j.tsf.2014.10.094

\bibitem{KOEHN.Köehler}
Köehn, D., Köehler, S., Toussaint, R., Ghani, I., Stollhofen, H.,
\textit{Scaling analysis, correlation length and compaction estimates of natural and simulated stylolites},
Journal of Structural Geology, 161, 104670, 2022. DOI: 10.1016/j.jsg.2022.104670'


\bibitem{Oliveira.Reis}
Oliveira, T.J., Aarão Reis, F.D.A.,
\textit{Simulating the initial growth of a deposit from colloidal suspensions},
Journal of Statistical Mechanics: Theory and Experiment, 2014(9), P09006. DOI: 10.1088/1742-5468/2014/09/P09006

\bibitem{Ghavami2025}
Feyzollah Ghavami Mirmahalle, S., Erfanifam, M., and Maleki, M.,
\textit{Dynamics of the surface growth resulted from deposition of free-falling spheres at the bottom of a Hele-Shaw cell},
Soft Matter, \textbf{21}, 8498 (2025).

\bibitem{RituparnoMandal}
Rituparno Mandal, Corneel Casert, Peter Sollich,
\textit{Robust prediction of force chains in jammed solids using graph neural networks},
Nature Communications, 2022.

\bibitem{Cheng}
Cheng, Z., Wang, J., Xiong, W.,
\textit{A machine learning-based strategy for experimentally estimating force chains of granular materials using X-ray micro-tomography},
Géotechnique, 2023. DOI: 10.1680/jgeot.21.00281


\bibitem{QU2021103046}
Tongming Qu, Shaocheng Di, Y.T. Feng, Min Wang, Tingting Zhao,
Towards data-driven constitutive modelling for granular materials via micromechanics-informed deep learning,
\textit{International Journal of Plasticity}, 144, 103046, 2021.


\bibitem{YE2022109437}
Ye, X., Ni, Y.-Q., Sajjadi, M., Wang, Y.-W., Lin, C.-S.,
\textit{Physics-guided, data-refined modeling of granular material-filled particle dampers by deep transfer learning},
Mechanical Systems and Signal Processing, 180, 109437, 2022. DOI: 10.1016/j.ymssp.2022.109437


\bibitem{Tianshu}
Song, T., Xia, H.,
\textit{Numerically stable neural network for simulating Kardar-Parisi-Zhang growth in the presence of uncorrelated and correlated noises},
Computer Physics Communications, 315, 109682, 2025. DOI: 10.1016/j.cpc.2025.109682


\bibitem{Goodfellow}
Goodfellow, I., Pouget-Abadie, J., Mirza, M., Xu, B., Warde-Farley, D., Ozair, S., Courville, A., Bengio, Y.,
\textit{Generative Adversarial Networks},
Advances in Neural Information Processing Systems, 2014, Vol. 3. DOI: 10.1145/3422622

\bibitem{Mirza}
Mirza, M., Osindero, S.,
\textit{Conditional Generative Adversarial Nets}, 2014.

\bibitem{Isola}
Isola, P., Zhu, J.-Y., Zhou, T., Efros, A.,
\textit{Image-to-Image Translation with Conditional Adversarial Networks},
CVPR, 2017, pp. 5967–5976. DOI: 10.1109/CVPR.2017.632

\bibitem{lammps}
Thompson, A.P., Aktulga, H.M., Berger, R., Bolintineanu, D.S., Brown, W.M., Crozier, P.S., In’t Veld, P.J., Kohlmeyer, A., Moore, S.G., Nguyen, T.D., et al.,
\textit{LAMMPS—a flexible simulation tool for particle-based materials modeling at the atomic, meso, and continuum scales},
Computer Physics Communications, 271, 108171, 2022

 
\end{thebibliography}
\end{document}